\documentclass[12pt]{article}
\usepackage{axodraw}

\newcommand{\beq}{\begin{equation}}
\newcommand{\eeq}{\end{equation}}
\newcommand{\beqa}{\begin{eqnarray}}
\newcommand{\eeqa}{\end{eqnarray}}
\newcommand{\beqar}{\begin{eqnarray*}}
\newcommand{\eeqar}{\end{eqnarray*}}

\newcommand{\al}{\alpha}
\newcommand{\be}{\beta}

\def\spa          {\ \ \ }
\def\non          {\nonumber}
\def\ha           {\mbox{$\frac{1}{2}$}}

\def\d        {\mbox{d}}
\def\spa          {\ \ \ }
\def\mand         {\spa\mbox{and}\spa}

\def\Tr           {\mbox{\rm Tr}\,}
\def\STr          {\mbox{\rm STr}\,}

\def\cd           {{\cdot}}
\def\ran          {\rangle}
\def\lan          {\langle}
\def\fsH	{H\!\!\!\!/\,}
\newcommand{\del}{\delta}

\newcommand{\eps}{\epsilon}
\newcommand{\ga}{\gamma}
\newcommand{\Ga}{\Gamma}

\newcommand{\h}{\eta}

\newcommand{\eg}{{\it e.g.,}\ }
\newcommand{\ie}{{\it i.e.,}\ }
\newcommand{\labell}[1]{\label{#1}} 
\newcommand{\reef}[1]{(\ref{#1})}
\newcommand\prt{\partial}

\newcommand\bD{\bar{D}}

\newcommand\bz{\bar{z}}

\begin{document}
\baselineskip 18pt%
\begin{titlepage}
\vspace*{1mm}%
\hfill%
\vspace*{15mm}%

\centerline{{\Large {\bf On Wess-Zumino terms of }}}\vspace*{3mm} \centerline{{\Large {\bf   Brane-Antibrane systems }}}
\vspace*{5mm}
\begin{center}
{ Mohammad R. Garousi$^{a,b}$ and Ehsan Hatefi$^{a}$}%

\vspace*{0.8cm}{ {${}^a{}$Department of Physics, Ferdowsi university,
P.O. Box 1436, Mashhad, Iran}}\\
{ {${}^b{}$ Institute for Studies in Theoretical Physics and Mathematics (IPM) \\
P.O.Box 19395-5531, Tehran, Iran}}\\
\vspace*{1.5cm}
\end{center}

\begin{center}{\bf Abstract}\end{center}
\begin{quote}
We calculate the disk level S-matrix element of one closed string RR field, two open string tachyons  and one gauge field in   type II superstring 
theory. An expansion for the S-matrix element has been found that its   four leading order terms   are reproduced exactly by  the symmetric trace tachyon DBI and the Wess-Zumino actions of D-brane-anti-D-brane systems. Using this consistency, we have  also found  the first higher derivative correction to the some of the  WZ terms.

\end{quote}
\end{titlepage}
\section{Introduction}
Study of unstable objects in string theory  might shed new light
in understanding properties of string theory in time-dependent
backgrounds \cite{Gutperle:2002ai,Sen:2002in,Sen:2002an,Sen:2002vv,Lambert:2003zr,Sen:2004nf}. Generally
speaking,  source of instability in these processes  is appearance
of some tachyonic  modes  in the spectrum of these 
objects. It  then makes sense to study them in a field
theory which includes those modes. In this regard, it has been
shown by A. Sen that an effective action of  Born-Infeld type
proposed in \cite{Sen:1999md,Garousi:2000tr,Bergshoeff:2000dq,Kluson:2000iy} can capture many properties
of the decay of non-BPS D$_p$-branes in string theory
\cite{Sen:2002in,Sen:2002an}. 

Recently, unstable objects have been used to study spontaneous chiral symmetry breaking  in  holographic model of QCD \cite{Casero:2007ae,Bergman:2007pm,Dhar:2007bz}. In these studies, flavor branes introduced by placing a set of parallel branes and antibranes on a background dual to a confining color theory \cite{Sakai:2004cn}. 
Detailed study of brane-antibrane system
 reveals   when brane separation is smaller than the
string length scale,  spectrum of this system has two tachyonic
modes \cite{Sen:1998ii}.  The
 effective action  should then include
these  modes because they are the most important ones
which rule  the dynamics of   the system.

The effective action of a $D_p\bD_p$-brane in Type IIA(B) theory should be  given by some extension of  the DBI action and the WZ terms which include the tachyon fields. The DBI part may be given by the projection of the effective action of two non-BPS $D_p$-branes in Type IIB(A) theory with $(-1)^{F_L}$ \cite{Garousi:2004rd}. We are interested in this paper in the appearance of tachyon, gauge field and the RR field in these actions. These fields appear in  the DBI part as the following \cite{Garousi:2007fn}:  
\beqa
S_{DBI}&=&-\int
d^{p+1}\sigma \Tr\left(V({\cal T})
\sqrt{-\det(\eta_{ab}
+2\pi\alpha'F_{ab}+2\pi\alpha'D_a{\cal T}D_b{\cal T})} \right)\,\,,\labell{nonab} \eeqa The trace in the above action 
should be completely symmetric between all  matrices
of the form $F_{ab},D_a{\cal T}$, and individual
${\cal T}$ of the tachyon potential.   These matrices  are
\beqa
F_{ab}=\pmatrix{F^{(1)}_{ab}&0\cr 
0&F^{(2)}_{ab}},\,\,
D_{a}{\cal T}=\pmatrix{0&D_aT\cr 
(D_aT)^*&0},\,\, {\cal T}=\pmatrix{0&T\cr 
T^*&0}\,\labell{M12} \eeqa 
where $F^{(i)}_{ab}=\prt_{a}A^{(i)}_{b}-\prt_{b}A^{(i)}_{a}$ and $D_{a}T=\prt_{a}T-i(A^{(1)}_a-A^{(2)}_a)T$. If one uses ordinary trace, instead, the above action reduces to the action proposed by A.Sen \cite{Sen:2003tm} after making  the kinetic term symmetric and performing the trace. This latter action is not consistent with S-matrix calculation.
The tachyon potential which is consistent with S-matrix element calculations
has the following expansion: \beqa
V(|T|)&=&1+\pi\alpha'm^2|T|^2+
\frac{1}{2}(\pi\alpha'm^2|T|^2)^2+\cdots
\non\eeqa  where $T_{p}$ is the
p-brane tension , $m^2$ is the mass squared of tachyon, \ie
$m^2=-1/(2\alpha')$. The above expansion is consistent with the
potential $V(|T|)=e^{\pi\alpha'm^2|T|^2}$ which is the tachyon
potential of BSFT ~\cite{Kutasov:2000aq}.

The terms of the above action which has
contribution to the S-matrix element of one gauge field and two
tachyons in which we are interested in this paper are the following \cite{Garousi:2007fn}: 
\beqa {\cal
L}_{DBI}&\!\!\!=\!\!\!&-T_p(2\pi\alpha')\left(m^2|T|^2+DT\cdot(DT)^{*}-\frac{\pi\alpha'}{2}
\left(F^{(1)}\cdot{F^{(1)}}+
F^{(2)}\cdot{F^{(2)}}\right)\right)+T_p(\pi\alpha')^3\nonumber\\
&&\times\left(\frac{2}{3}DT\cdot(DT)^{*}\left(F^{(1)}\cdot{F^{(1)}}+F^{(1)}\cdot{F^{(2)}}+F^{(2)}\cdot{F^{(2)}}\right)\right.\labell{exp1}\\
&&\left.+\frac{2m^2}{3}|\tau|^2\left(F^{(1)}\cdot{F^{(1)}}+F^{(1)}\cdot{F^{(2)}}+F^{(2)}\cdot{F^{(2)}}\right)\right.\nonumber\\
&&-\left.\frac{4}{3}\left((D^{\mu}T)^*D_{\beta}T+D^{\mu}T(D_{\beta}T)^*\right)\left({F^{(1)}}^{\mu\alpha}F^{(1)}_{\alpha\beta}+{F^{(1)}}^{\mu\alpha}F^{(2)}_{\alpha\beta}+{F^{(2)}}^{\mu\alpha}F^{(2)}_{\alpha\beta}\right)\right)
\nonumber
\eeqa
 Note that if one uses the on-shell value for the tachyon mass, \ie $m^2=-1/(2\alpha')$,  the above terms would not be ordered in terms of power of $\alpha'$. 

The  WZ 
term describing the  coupling of RR field to gauge field of brane-anti-brane is given by~\cite{Douglas:1995bn,Li:1995pq}
\beqa
S = \mu_p\int_{\Sigma_{(p+1)}} C \wedge  \left(e^{i2\pi\alpha'F^{(1)}}-e^{i2\pi\alpha'F^{(2)}}\right)\ ,
\labell{eqn.wz}
\eeqa
where $\Sigma_{(p+1)}$ is the world volume and $\mu_p$ is the RR charge of the branes. In above equation, $C$ is a formal sum of the RR potentials $C=\sum_n(-i)^{\frac{p-m+1}{2}}C_m$. Note that the factors of $i$ disappear in  each term of \reef{eqn.wz}.
The inclusion of the tachyon fields into this action has been proposed in \cite{Kennedy:1999nn,Kraus:2000nj,Takayanagi:2000rz} using the superconnection of noncommutative geometry ~\cite{quil,berl,Roepstorff:1998vh}\beqa
S_{WZ}&=&\mu_p \int_{\Sigma_{(p+1)}} C \wedge \STr e^{i2\pi\alpha'\cal F}\labell{WZ}\eeqa 
where the curvature of the superconnection is defined as:
\beqa {\cal F}&=&d{\cal A}-i{\cal A}\wedge\cal A\eeqa
the superconnection is \begin{displaymath}
i{\cal A} = \left(
\begin{array}{cc}
  iA^{(1)} & \beta T^* \\ \beta T &   iA^{(2)} 
\end{array}
\right) \ ,
\non\end{displaymath}
where $\beta$ is a normalization constant with dimension $1/\sqrt{\alpha'}$ which we shall find it later, and  a ``supertrace'' is defined by
\begin{displaymath}
\STr \left( \begin{array}{cc} A&B\\C&D \end{array} \right)
= \Tr A - \Tr D \ .
\non\end{displaymath} Using the multiplication rule of the supermatrices \cite{Kraus:2000nj}
\beqa
 \left
( \begin{array}{cc}A&B\\C&D\end{array} 
\right)\cdot \left( \begin{array}{cc}A'&B'\\C'&D'\end{array} 
\right)\,=\,\left( \begin{array}{cc}AA'+(-)^{c'}BC'&AB'+(-)^{d'}BD'\\DC'+(-)^{a'}CA'&DD'+(-)^{b'}CB'\end{array} 
\right)\eeqa where $x'$ is 0 if $X$ is an even form or 1 if $X$ is an odd form, one finds that the curvature  is
\begin{displaymath}
i{\cal F} = \left(
\begin{array}{cc}
iF^{(1)} -\beta^2 |T|^2 & \beta (DT)^* \\
\beta DT & iF^{(2)} -\beta^2|T|^2 
\end{array}
\right) \ ,
\non\end{displaymath}
where $F^{(i)}=\frac{1}{2}F^{(i)}_{ab}dx^{a}\wedge dx^{b}$ and $DT=[\partial_a T-i(A^{(1)}_{a}-A^{(2)}_{a})T]dx^{a}$. The WZ action \reef{WZ} has the following terms:
\beqa
C\wedge \STr i{\cal F}&\!\!\!\!=\!\!\!&C_{p-1}\wedge(F^{(1)}-F^{(2)})\labell{exp2}\\
C\wedge \STr i{\cal F}\wedge i{\cal F}&\!\!\!\!=\!\!\!\!&C_{p-3}\wedge \left\{F^{(1)}\wedge F^{(1)}-
F^{(2)}\wedge F^{(2)}\right\}\nonumber\\
&& +C_{p-1}\wedge\left\{-2\beta^2|T|^2(F^{(1)}-F^{(2)})+2i\beta^2 DT\wedge(DT)^*\right\}\nonumber\\
C\wedge \STr i{\cal F}\wedge i{\cal F}\wedge i{\cal F}&\!\!\!\!=\!\!\!\!&
C_{p-5}\wedge \left\{F^{(1)}\wedge F^{(1)}\wedge F^{(1)}-
F^{(2)}\wedge F^{(2)}\wedge F^{(2)}\right\}\nonumber\\
&&+C_{p-3}\left\{-3\beta^2|T|^2(F^{(1)}\wedge F^{(1)}-F^{(2)}\wedge F^{(2)})\right.\nonumber\\
&&\left.\qquad \qquad+3i\beta^2(F^{(1)}+F^{(2)})\wedge DT\wedge(DT)^*\right\}\nonumber\\
&&+C_{p-1}\left\{3\beta^4|T|^4\wedge(F^{(1)}-F^{(2)})-6i\beta^4|T|^2DT\wedge (DT)^*\right\}\nonumber
\eeqa
The appearance of $C_{p-1}\wedge dT\wedge dT^*$ has been checked in \cite{Kennedy:1999nn} by studying the disk level S-matrix element of one RR field and two tachyons. In the present  paper we will check, among other things, the appearance of $C_{p-1}\wedge DT\wedge (DT)^*$ and $C_{p-1}\wedge |T|^2(F^{(1)}-F^{(2)})$ terms and fix their coefficients using the  S-matrix element of one RR field, two tachyons and one gauge field. The coupling of one RR field, two tachyons and one gauge field in the above terms  can be combined into the following form:
\beqa
&\mu_p(2\pi\alpha')^2(-\beta^2) \int_{\Sigma_{(p+1)}} C_{(p-1)}\wedge \left\{d(A^{(1)}-A^{(2)})TT^*-(A^{(1)}-A^{(2)})d(TT^*)\right\}
&\nonumber\\
&=\mu_p(2\pi\alpha')^2(-\beta^2)\int_{\Sigma_{p+1}}H_{(p)}\wedge (A^{(1)}-A^{(2)})TT^*&\labell{contact}\eeqa
This combination actually appears naturally in the S-matrix element in the string theory side.

An outline of the rest of paper is as follows. In the next section,  we review the calculate of the S-matrix element of one RR and two tachyons \cite{Kennedy:1999nn}. The low energy  expansion of this amplitude produces one massless pole and infinite number of contact terms. The massless pole is the one reproduce by the above field theories, and its first contact term by $C_{p-1}\wedge dT\wedge dT^*$. This fixes the normalization of tachyon in WZ action with respect to the tachyon in the DBI part. In section 3, we calculate the S-matrix element of one RR, two tachyons and one gauge field. We find an expansion for the amplitude whose leading order terms are fully consistent with the above field theories.


\section{The $T-T-C$ amplitude}

The three-point amplitude between one RR field and two  tachyons  has been studied in \cite{Kennedy:1999nn} where a non-zero coupling $C\wedge dT\wedge dT^*$ has been found. To fix the coefficient of this term  we reexamine this amplitude in this section.  The tachyon vertex operator corresponds to the real  components of the complex tachyon, \ie
\beqa
T&=&\frac{1}{\sqrt{2}}(T_1+iT_2)\eeqa
 The three point amplitude between one RR and two  tachyons is given by the following correlation functions:
\begin{eqnarray}
{\cal A}^{T,T,RR} & \sim & \int dxdyd^2w
 \lan V_{T}^{(0)}(x)
V_{T}^{(-1)}(y)
V_{RR}^{(-1)}(w,\bar{w})\ran\labell{cor1}\eeqa
where the vertex operators are
\beqa
V_{T}^{(0)}(x) &=&2ik\cd\psi(x) e^{2ik\cd X(x)}
\nonumber\\
V_{T}^{(-1)}(y) &=& e^{-\phi(y)} e^{2ik'\cd X(y)}\labell{vertex1}\\
V_{RR}^{(-1)}(w,\bar{w})&=&(P_{-}\fsH_{(n)}M_p)^{\al\be}e^{-\phi(w)/2} S_{\al}(w)e^{ip\cd X(w)}e^{-\phi(\bar{w})/2} S_{\be}(\bar{w}) e^{ip\cd D \cd X(\bar{w})}\nonumber
\eeqa
The tachyon vertex operator corresponds to  either  $T_1$ or $T_2$. The projector in the RR vertex operator is $P_{-} = \ha (1-\ga^{11})$  and
\begin{displaymath}
\fsH_{(n)} = \frac{a
_n}{n!}H_{\mu_{1}\ldots\mu_{n}}\ga^{\mu_{1}}\ldots
\ga^{\mu_{n}}
\ ,
\non\end{displaymath}
where $n=2,4$ for type IIA and $n=1,3,5$ for type IIB. $a_n=i$ for IIA and $a_n=1$ for IIB theory. The spinorial
indices are raised with the charge conjugation matrix, eg
$(P_{-}\fsH_{(n)})^{\al\be} =
C^{\al\del}(P_{-}\fsH_{(n)})_{\del}{}^{\be}$ (further conventions and
notations for spinors can be found in appendix~B of~\cite{Garousi:1996ad}).  The
RR bosons are massless so $p^{2}=0$ and for the tachyons $k^2=k'^2=1/4$. In the string theory calculation we always  set $\alpha'=2$.  The world-sheet  fields has been
extended to the entire complex plane.  That is,  we have replaced 
\begin{displaymath}
\tilde{X}^{\mu}(\bar{w}) \rightarrow D^{\mu}_{\nu}X^{\nu}(\bar{w}) \ ,
\spa
\tilde{\psi}^{\mu}(\bar{w}) \rightarrow
D^{\mu}_{\nu}\psi^{\nu}(\bar{w}) \ ,
\spa
\tilde{\phi}(\bar{w}) \rightarrow \phi(\bar{w})\,, \mand
\tilde{S}_{\al}(\bar{w}) \rightarrow M_{\al}{}^{\be}{S}_{\be}(\bar{w})
 \ ,
\non\end{displaymath}
where
\begin{displaymath}
D = \left( \begin{array}{cc}
1_{p+1} & 0 \\
0 & -1_{9-p}
\end{array}
\right) \ ,\,\, \mand 
M_p = \left\{\begin{array}{cc}\frac{\pm i}{(p+1)!}\ga^{a_{0}}\ga^{a_{1}}\ldots \ga^{a_{p}}
\eps_{a_{0}\ldots a_{p}}\,\,\,\,{\rm for\, p \,even}\\ \frac{\pm 1}{(p+1)!}\ga^{a_{0}}\ga^{a_{1}}\ldots \ga^{a_{p}}\ga_{11}
\eps_{a_{0}\ldots a_{p}} \,\,\,\,{\rm for\, p \,odd}\end{array}\right. 
\non\end{displaymath}
Using these replacements, one finds the standard propagators for the world-sheet fields $X^{\mu}\,, \phi$, \ie
\begin{eqnarray}
\lan X^{\mu}(z)X^{\nu}(w)\ran & = & -\eta^{\mu\nu}\log(z-w) \ , \non \\
\lan\phi(z)\phi(w)\ran & = & -\log(z-w) \ .
\labell{prop}\end{eqnarray}
One also needs the correlation function between two spin operators and one $\psi$.
The correlation function involving an arbitrary number of $\psi$'s 
and 
two $S$'s is obtained using the following Wick-like rule~\cite{Liu:2001qa}:
 \beqa
 \lan\psi^{\mu_1}
(y_1)...
\psi^{\mu_n}(y_n)S_{\al}(z)S_{\be}(\bz)\ran&\!\!\!\!=\!\!\!\!
&\frac{1}{2^{n/2}}
\frac{(z-\bz)^{n/2-5/4}}
{|y_1-z|...|y_n-z|}\left[(\Gamma^{\mu_n...\mu_1}
C^{-1})_{\al\be}\right.\nonumber\\&&+
\lan\psi^{\mu_1}(y_1)\psi^{\mu_2}(y_2)\ran(\Gamma^{\mu_n...\mu_3}
C^{-1})_{\al\be}
\pm perms\nonumber\\&&+\lan\psi^{\mu_1}(y_1)\psi^{\mu_2}(y_2)\ran
\lan\psi^{\mu_3}(y_3)\psi^{\mu_4}(y_4)\ran(\Gamma^{\mu_n...\mu_5}
C^{-1})_{\al\be}\nonumber\\&&\left.
\pm perms+\cdots\right]\labell{wicklike}\eeqa where  dots means  sum  over all possible contractions. In above equation, $\Gamma^{\mu_{n}...\mu_{1}}$ is the totally antisymmetric combination of the gamma matrices and  the Wick-like contraction, for  real $y_i$, is given by \beqar\lan\psi^{\mu}(y_{1})\psi^{\nu}(y_{2})\ran &=&2\eta^{\mu\nu}{\frac {Re[(y_{1}-z)(y_{2}-\bz)]}{(y_{1}-y_{2})(z-\bz)}}\eeqar 
The number of $\psi$ in the correlators  \reef{cor1} is one. Using the above formula for one $\psi$ and performing the other correlators using \reef{prop}, one finds that the integrand  is invariant under $SL(2,R)$ transformation. Gauge fixing this symmetry by 
 fixing  the position of vertex operators at  $(x,y,w,\bar{w}) = (x,-x,i,-i)$, one finds \cite{Kennedy:1999nn} 
\beqa
{\cal A}^{T,T,RR} & \sim & \int_{-\infty}^{\infty} \d x
\left( \frac{(1+x^{2})^{2}}{16 x^{2}}\right)^{\ha + u} \frac{2}{1+x^{2}} \Tr
(P_{-}\fsH_{(n)}M_p\ga^{a})k_{a} \ , \non \\
&=&\left(\frac{i\mu_p}{4}\right)2\pi \frac{\Ga[-2u]}{\Ga[\ha-u]^2}
 \Tr (P_{-}\fsH_{(n)}M_p\ga^{a})k_{a} \labell{amp33}\ .
\eeqa
where $u = -(k+k')^2$ and conservation of momentum along the world volume of brane, \ie $k^{a} + k'{}^{a} + p^{a} =0$, has been used. We have also normalized the amplitude by $i\mu_p/4$.  The trace is zero for $p\neq n$, and for $n=p$ it is
\begin{displaymath}
\Tr \left( \fsH_{(n)}M_p\ga^{a}\right)
= \pm\frac{ 32}{p!}H_{a_{0}\ldots a_{p-1}}
 \eps^{a_{0}\ldots a_{p-1}a} \ .
\non\end{displaymath}
We are going to compare string theory S-matrix elements with field theory S-matrix element including their coefficients, however, we are not interested in fixing the overall sign of the amplitudes. Hence, in above and in the rest of equations in this paper, we have payed   no attention to the sign of  equations. The trace in \reef{amp33} containing the factor of $\ga^{11}$ ensures the following
results also hold for $p>3$ with $H_{(n)} \equiv \ast H_{(10-n)}$ for
$n\geq 5$. 

If one  replaces $k_a$ in \reef{amp33} with $-k'_a-p_a$ and uses using the conservation of momentum,  one will find  that the $p_a$ term vanishes using the totally antisymmetric property of $\eps^{a_{0}\ldots a_{p-1}a}$. Hence the amplitude \reef{amp33} is antisymmetric under interchanging $1\leftrightarrow 2$. This indicates that the three point amplitude between one RR and two $T_1$ or two $T_2$ is zero. 
\begin{center}
\begin{picture}
(600,100)(0,0)
\Line(25,105)(75,70)\Text(50,105)[]{$T_{1}$}
\Line(25,35)(75,70)\Text(50,39)[]{$T_{2}$}
\Photon(75,70)(125,70){4}{7.5}\Text(100,88)[]{$A$}
\Gluon(125,70)(175,105){4.20}{5}\Text(145,105)[]{$C_{p-1}$}

\SetColor{Black}
\Vertex(75,70){1.5} \Vertex(125,70){1.5}
\Text(105,20)[]{(a)}
\Line(295,105)(345,70)\Text(320,105)[]{$T_{1}$}
\Line(295,35)(345,70)\Text(310,35)[]{$T_2$}
\Gluon(345,70)(395,105){4.20}{5}\Text(365,105)[]{$C_{p-1}$}
\Vertex(345,70){1.5}
\Text(345,20)[]{(b)}

\end{picture}\\ {\sl Figure 1 : The Feynman diagrams corresponding to the amplitudes \reef{amp2} and \reef{amp44}.}  
\end{center}

The  effective Lagrangian of massless and tachyonic
particles, \reef{exp1} and \reef{exp2},  produces the following massless pole:
\beqa
{\cal A}&=&V_a(C_{p-1},A)G_{ab}(A)V_b(A,T_1,T_2)\labell{amp2}\eeqa
where the gauge field in the off-shell line is $A^{(1)}$ and $A^{(2)}$. The propagator and vertexes  are
\beqa
G_{ab}(A) &=&\frac{i\delta_{ab}}{(2\pi\alpha')^2 T_p
\left(-p_a^2\right)}\nonumber\\
V_b(A^{(1)},T_1,T_2)&=&iT_p(2\pi\alpha')(k_b-k'_b)\nonumber\\
V_b(A^{(2)},T_1,T_2)&=&-iT_p(2\pi\alpha')(k_b-k'_b)\nonumber\\
V_a(C_{p-1},A^{(1)})&=&i\mu_p(2\pi\alpha')\frac{1}{p!}\epsilon_{a_0\cdots a_{p-1}a}H^{a_0\cdots a_{p-1}}\nonumber\\
V_a(C_{p-1},A^{(2)})&=&-i\mu_p(2\pi\alpha')\frac{1}{p!}\epsilon_{a_0\cdots a_{p-1}a}H^{a_0\cdots a_{p-1}}\eeqa
Replacing them in  \reef{amp2}, one finds 
\beqa
{\cal A}&=&4i\mu_p\frac{1}{p!u}\epsilon_{a_0\cdots a_{p-1}a}H^{a_0\cdots a_{p-1}}k^a\labell{amp444}\eeqa
The field theory \reef{exp2} has also the following contact term:
\beqa
{\cal A}_c&=&i\mu_p(2\pi\alpha')^2\beta^2\frac{1}{p!}\epsilon_{a_0\cdots a_{p-1}a}H^{a_0\cdots a_{p-1}}k^a\labell{amp44}\eeqa

The  massless pole of field theory \reef{amp444}  can be obtained from the S-matrix element \reef{amp33} by expanding it at low energy ($u = -p_{a}^{2} \rightarrow 0$). The
prefactor of \reef{amp33} has the expansion 
\beqa
2\pi\frac{\Ga[-2u]}{\Ga[\ha-u]^2}
 &=&  \frac{-1}{
 u} + 4\ln(2) +\left(\frac{\pi^2}{6}-8\ln(2)^2\right)u+ O(u^2)
\ .\labell{taylor}
\eeqa
The first term is exactly the field theory massless pole \reef{amp444}.  
The second term should be  the contact term in \reef{amp44}.
 This fixes the normalization constant $\beta$ to be \beqa
\beta&=&\frac{1}{\pi} \sqrt{\frac{2\ln(2)}{\alpha'}}\eeqa
Note that in the string theory calculations we have set $\alpha'=2$, so to compare the string theory amplitudes with the corresponding amplitudes in field theory one has to set $\alpha'=2$ in the field theory side too.

Since the expansion \reef{taylor} is in terms of the powers of $p_a^2$, the other  terms in \reef{taylor}  correspond to the higher derivative corrections of  the WZ action. For example, it is easy to check that the following  higher derivative term  reproduces the third term in \reef{taylor}:
\beqa
i(\alpha')^2\mu_p\left(\frac{\pi^2}{6}-8\ln(2)^2\right) C_{p-1}\wedge D ^aD_a(DT\wedge DT^*)\labell{hderv}
\eeqa
On the other hand, one may write $u=-p_a^2=-1/2-2k\cdot k'$ using the on-shell condition $k^2=k'^2=1/4$. Then one may  conclude  that because of the $-1/2$ term, the last term in \reef{taylor} does not correspond to the higher derivative of the tachyons. As we mentioned, it is obvious  that the $O(u)$ terms in the expansion \reef{taylor} correspond to the higher derivative of the tachyons, however, for other S-matrix elements, \eg the S-matrix element of four tachyons, it is hard to prove it. It is speculated though  that  the non-leading terms of the expansion of any S-matrix element of tachyons  correspond to the higher derivative of the tachyon field \cite{Garousi:2002wq}. 
 
It is interesting to compare the above higher derivative term with the higher derivative term of one RR and two gauge fields. The string theory S-matrix element of one RR and two gauge fields is given by \cite{Hashimoto:1996kf,Garousi:1998fg}
\beqa
{\cal A}&\sim&2\frac{\Gamma[-2u]}{\Gamma[1-u]^2}K\eeqa
where $K$ is the kinematic factor. Expansion of the prefactor at low energy is 
\beqa
2\frac{\Ga[-2u]}{\Ga[1-u]^2}
 &=&  \frac{-1}{
 u}  -\left(\frac{\pi^2}{6}\right)u+ O(u^2)
\ .\labell{taylor1}
\eeqa
In this case actually there is no massless pole at $u=0$ as the  kinematic factor  provides a compensating factor of $u$. The amplitude  has the following expansion:
\beqa
{\cal A}&=&i\frac{(4\pi)^2\mu_p}{4(p-3)!}f^{a_0a_1}{f'}^{a_2a_3}\varepsilon^{a_4\cdots a_p}\epsilon_{a_0\cdots a_p}\left(1+(\frac{\pi^2}{6})u^2+ O(u^3)\right)\delta_{p,n+2}
\eeqa
where $f_{ab}=i(k_a\xi_b-k_b\xi_a)$, ${f'}_{ab}=i(k'_a\xi'_b-k'_b\xi'_a)$ and $\varepsilon$ is the polarization of the RR potential. The first term is reproduced by the couplings  in the ZW terms \reef{exp2}, and the second term by the following higher derivative term
\beqa
\frac{(\pi\alpha')^2}{6}\frac{\mu_p}{2!}(2\pi\alpha')^2C_{p-3}\wedge \left(\partial^a\partial^b F^{(1)}\wedge \partial_a\partial_b F^{(1)}-\partial^a\partial^b F^{(2)}\wedge \partial_a\partial_b F^{(2)}\right)\labell{highaa}\eeqa
While the leading higher derivative term in \reef{hderv} has two extra derivatives with respect to the corresponding coupling in the WZ terms, in above coupling there are four extra derivatives with respect to the corresponding coupling in \reef{exp2}.

\section{The $T-T-A-C$ amplitude}
The S-matrix element of  one RR field,  two tachyons   and 
one gauge field is given by the following correlation function:
\begin{eqnarray}
{\cal A}^{ATTC} & \sim & \int dx_{1}dx_{2}dx_{3}dzd\bar{z}\,
  \lan V_A^{(-1)}{(x_{1})}V_{T}^{(0)}{(x_{2})}
V_{T}^{(0)}{(x_{3})}
V_{RR}^{(-1)}(z,\bar{z})\ran\eeqa
where we have chosen the vertex operators according to the rule that the
total superghost number must be $-2$ . The tachyon and RR vertex operators are given in \reef{vertex1} and the gauge field vertex operator in (-1)-picture is 
\beqa
V_A^{(-1)}&=&{\xi}.{\psi(x_{1})}e^{-\phi(x_1)}\,e^{2ik_{1}.X(x_{1})}\labell{vertex2}\eeqa
where $\xi_a$ is the gauge field polarization. 
Introducing $x_{4}\equiv\ z=x+iy$ and $x_{5}\equiv\bz=x-iy$,  the scattering amplitude reduces to the following 
 correlators:\beqar {\cal A}^{ATT^*C}&\sim& \int 
 dx_{1}\cdots dx_{5}\,
(P_{-}\fsH_{(n)}M_p)^{\al\be}\xi_a<:e^{-1/2\phi(x_4)}:
e^{-1/2\phi(x_5)}:e^{-\phi(x_1)}:>\non \\
&&\times {<:e^{2ik_1.X(x_1)}:e^{2ik_2.X(x_2)}
:e^{2ik_3.X(x_3)}:e^{ip.X(x_4)}:e^{ip.D.X(x_5)}:>}
\  \non \\&&\times {<:S_{\al}(x_4):S_{\be}(x_5)
:{\psi^a(x_1)}
:2k_2.{\psi(x_2)}:2k_3.{\psi(x_3)}:>}\eeqar
The correlators in the first and the second lines can be calculated using the propagators in \reef{prop}, and the correlator in the last line can read from \reef{wicklike}. The result is 
\begin{eqnarray}
{\cal A}^{ATTC}&\!\!\!\!\sim\!\!\!\!\!&\sqrt{2}\int dx_{1}\cdots dx_{5}\,k_{2b}k_{3c}
{\xi_a}(x_{34}x_{35})^{-1/2+2k_{3}.p}{(x_{14}x_{15})^{-1+2k_{1}.p}}
(x_{24}x_{25})^{-1/2+2k_{2}.p}
\nonumber\\
&& {x_{45}^{p.D.p}}{{x_{23}^{4k_2.k_3}x_{12}^
{4k_1.k_2}}{x_{13}^{4k_1.k_3}}}(P_{-}\fsH_{(n)}M_p)^{\al\be}
\bigg\{(\Gamma^{cba}C^{-1})_{\al\be}+2{\h^{ab}}{\frac{Re[x_{14}x_{25}]}
{x_{12}x_{45}}}(\Gamma^{c}C^{-1})_{\al\be}
\nonumber\\
&& -2{\h^{ac}}{\frac{Re[x_{14}x_{35}]}
{x_{13}x_{45}}}
(\Gamma^{b}C^{-1})_{\al\be}+
 2{\h^{bc}}{\frac{Re[x_{24}x_{35}]}{x_{23}x_{45}}}(\Gamma^{a}C^{-1})
 _{\al\be}\bigg\}\non\end{eqnarray}
where ${x_{ij}}\equiv\ x_{i}-x_{j}$. One can show that the integrand is invariant under
SL(2,R) transformation. Gauge fix  this  symmetry by fixing \beqar
 x_{1}&=&0 ,\qquad x_{2}=1,\qquad x_{3}\rightarrow \infty,
 \qquad dx_1dx_2dx_3\rightarrow x_3^{2}
 \eeqar One  finds 
 \beqa {\cal A}^{ATTC}&\sim&\sqrt{2}\int
dx_4dx_5\,k_{2b}k_{3c}{\xi_a}{x_{45}^
{-2(t+s+u)-1}}{|x_{4}|^{2t+2s-1}}{|1-x_{4}|^{2u+2t-1/2}}
\nonumber\\&&\times
 \bigg\{(\Gamma^{cba}C^{-1})_{\al\be}
-2{\h^{ab}}{\frac{-x+x_{4}x_{5}}{x_{45}}}(\Gamma^{c}C^{-1})_{\al\be}
-2{\h^{ac}}
{\frac{x}{x_{45}}}(\Gamma^{b}C^{-1})_{\al\be}\nonumber\\&&
+2{\h^{bc}}{\frac{x-1}{x_{45}}}(\Gamma^{a}C^{-1})_{\al\be}\bigg\}
(P_{-}\fsH_{(n)}M_p)^
 {\al\be}\nonumber\eeqa 
where we have also introduced   the Mandelstam variables \beqar
s&=&-(k_1+k_3)^2,\qquad t=-(k_1+k_2)^2,\qquad u=-(k_2+k_3)^2
\qquad\eeqar
 and used the conservation of momentum along the brane, \ie $k_{1}^{a}+k_{2}^{a}
+k_{3}^{a}+p^{a}=0$. 

Using the fact that $M_p$, $\fsH_{(n)}$, and $\Gamma^{cba}$ are totally antisymmetric combinations of the Gamma matrices, one realizes that the first term is non-zero only for $p=n+2$, and the last three terms are non-zero only for $p=n$. The integral in the above equation can be written in terms of the Gamma functions, using the following identity \cite{Fotopoulos:2001pt}:
\beqa 
 \int d^2 \!z |1-z|^{a} |z|^{b} (z - \bar{z})^{c}
(z + \bar{z})^{d}&\!\!\!\!=\!\!\!&
(2 \imath)^{c} 2^d \,  \pi \frac{ \Gamma( 1+ d +
\frac{b+c}{2})\Gamma( 1+ \frac{a+c}{2})\Gamma( -1-
\frac{a+b+c}{2})\Gamma( \frac{1+c}{2})}{
\Gamma(-\frac{a}{2})\Gamma(-\frac{b}{2})\Gamma(2+c+d+
\frac{a+b}{2})}\nonumber
\eeqa
for $ d= 0,1$ and arbitrary $a,b,c$. The region of integration is the upper half complex plane, as in our case. Using this integral, one finds
\beqa
{\cal A}^{ATTC}&=&\frac{i\mu_p}{2\sqrt{2\pi}}\left[\Tr\bigg((P_{-}\fsH_{(n)}M_p)
(k_3.\ga)(k_2.\ga)(\xi.\ga)\bigg)I\delta_{p,n+2}
+\Tr\bigg((P_{-}\fsH_{(n)}M_p)
\ga^{a}\bigg)J\delta_{p,n}\right. \nonumber\\&&\left.\times
\bigg\{k_{2a}(t+1/4)(2\xi.k_{3})
+k_{3a}(s+1/4)(2\xi.k_{2})-\xi_a(s+1/4)(t+1/4)\bigg\}\right]\labell{gen}\eeqa
where we have normalized the amplitude by $i\mu_p/2\sqrt{2\pi}$. In above equation,  $I,\,J$ are :
 \beqa I&=&2^{1/2}(2)^{-2(t+s+u)-1}\pi{\frac{\Gamma(-u)
 \Gamma(-s+1/4)\Gamma(-t+1/4)\Gamma(-t-s-u)}{\Gamma(-u-t+1/4)
\Gamma(-t-s+1/2)\Gamma(-s-u+1/4)}}\nonumber\eeqa
\beqa J&=&2^{1/2}(2)^{-2(t+s+u+1)}\pi{\frac{\Gamma(-u+1/2)
\Gamma(-s-1/4)\Gamma(-t-1/4)\Gamma(-t-s-u-1/2)}
{\Gamma(-u-t+1/4)\Gamma(-t-s+1/2)\Gamma(-s-u+1/4)}}\nonumber\eeqa 
The  traces are: \beqa
\Tr\bigg(\fsH_{(n)}M_p
(k_3.\ga)(k_2.\ga)(\xi.\ga)\bigg)\delta_{p,n+2}&=&\pm\frac{32}{n!}\eps^{a_{0}\cdots a_{p}}H_{a_{0}\cdots a_{p-3}}
k_{3a_{p-2}}k_{2a_{p-1}}\xi_{a_p}\delta_{p,n+2}\nonumber\\
\Tr\bigg(\fsH_{(n)}M_p
\ga^a\bigg)\delta_{p,n}&=&\pm\frac{32}{n!}\eps^{a_{0}\cdots a_{p-1}a}H_{a_{0}\cdots a_{p-1}}
\delta_{p,n}
\eeqa
Examining  the poles of the Gamma functions, one realizes that for the case that $p=n+2$, the amplitude has massless pole and infinite tower of massive poles. Whereas for $p=n$ case, there are tachyon, massless, and infinite tower of massive poles. Now the non trivial question is how to expand this amplitude such that its leading terms correspond to the effective actions \reef{exp1} and \reef{exp2}, and its non leading terms correspond to the higher derivative terms? Let us study each case separately. 

\subsection{$p=n+2$ case}

For $p=n+2$, the amplitude is antisymmetric under interchanging $ 2 \leftrightarrow 3$, hence the four-point function between one RR, one gauge field and two $T_1$ or two $T_2$ is zero. The electric part of the amplitude for one RR, one gauge field, one $T_1$ and one $T_2$ is given by 
\beqa
{\cal A}^{AT_1T_2C}&=&\pm\frac{8i\mu_p}{\sqrt{2\pi}(p-2)!}\left[ \eps^{a_{0}\cdots a_{p}}H_{a_{0}\cdots a_{p-3}}
k_{3a_{p-2}}k_{2a_{p-1}}\xi_{a_p}\right]I\labell{pn2}\eeqa 
Note that the amplitude satisfies the Ward identity, \ie the amplitude vanishes under replacement $\xi^a\rightarrow k_1^a$. 
\begin{center}
\begin{picture}
(600,100)(0,0)
\Line(25,105)(75,70)\Text(50,105)[]{$T_{1}$}
\Line(25,35)(75,70)\Text(50,39)[]{$T_{2}$}
\Photon(75,70)(125,70){4}{7.5}\Text(105,88)[]{$A^{(1)}$}
\Gluon(125,70)(175,105){4.20}{5}\Text(145,105)[]{$C_{p-3}$}
\Photon(125,70)(175,35){4}{7.5}\Text(150,39)[]{$A^{(1)}$}
\SetColor{Black}
\Vertex(75,70){1.5} \Vertex(125,70){1.5}
\Text(105,20)[]{(a)}
\Line(295,105)(345,70)\Text(320,105)[]{$T_{1}$}
\Line(295,70)(345,70)\Text(315,80)[]{$T_2$}
\Photon(295,35)(345,70){4}{7.5}\Text(315,35)[]{$A^{(1)}$}
\Gluon(345,70)(395,105){4.20}{5}\Text(365,105)[]{$C_{p-3}$}
\Vertex(345,70){1.5}
\Text(345,20)[]{(b)}

\end{picture}\\ {\sl Figure 2 : The Feynman diagrams corresponding to the amplitudes \reef{amp3} and \reef{amp41}.} 
\end{center}

The  effective Lagrangian of the massless field and tachyon, \reef{exp1} and \reef{exp2},  produces 
 the following massless pole for $p=n+2$:
\beqa
{\cal A}&=&V_a(C_{p-3},A^{(1)},A^{(1)})G_{ab}(A^{(1)})V_b(A^{(1)},T_1,T_2)\labell{amp3}\eeqa
where
\beqa
G_{ab}(A^{(1)}) &=&\frac{i\delta_{ab}}{(2\pi\alpha')^2 T_p
\left(u\right)}\nonumber\\
V_b(A^{(1)},T_1,T_2)&=&T_p(2\pi\alpha')(k_2-k_3)_b\nonumber\\
V_a(C_{p-3},A^{(1)},A^{(1)})&=&\mu_p(2\pi\alpha')^2\frac{1}{(p-2)!}\epsilon_{a_0\cdots a_{p-1}a}H^{a_0\cdots a_{p-3}}k_1^{a_{p-2}}\xi^{a_{p-1}}\eeqa
The amplitude \reef{amp3} becomes
\beqa
{\cal A}&=&\mu_p(2\pi\alpha')\frac{2i}{(p-2)!u}\epsilon_{a_0\cdots a_{p-1}a}H^{a_0\cdots a_{p-3}}k_2^{a_{p-2}}k_3^{a_{p-1}}\xi^a\labell{amp40}\eeqa
The WZ action  \reef{exp2} has also the following contact term:
\beqa
{\cal A}_c&=&\mu_p\beta^2(2\pi\alpha')^3\frac{i}{2!(p-2)!} \epsilon_{a_0\cdots a_{p-1}a}H^{a_0\cdots a_{p-3}}k_2^{a_{p-2}}k_3^{a_{p-1}}\xi^a\labell{amp41}\eeqa
The  massless pole \reef{amp40} and the contact term \reef{amp41}  can be exactly reproduced by the string theory amplitude \reef{gen} if one expands it around 
\beqa t\rightarrow{-1/4},\qquad s\rightarrow{-1/4},\qquad
u\rightarrow{0} \labell{point}\eeqa  
In fact expansion of $I$ around this point is 
\beqa
I&=&
\pi\sqrt{2\pi}\left(\frac{-1}{u}
+4\ln(2)+\right.\nonumber\\
&&\left.+\left(\frac{\pi^2}{6}-8\ln(2)^2\right)u-\frac{\pi^2}{6}\frac{(s+t+1/2)^2}{u}+\cdots\right)\labell{line}\eeqa
replacing it in \reef{pn2}, one finds that the first term is reproduced by \reef{amp40} and the second term by \reef{amp41}.

Now what about the other terms of \reef{line}? A natural extension of the higher derivative term \reef{hderv} to $C_{p-3}$ which is similar to the extension of $C_{p-1}\wedge(DT\wedge DT^*)$ to $C_{p-3}$ in \reef{exp2} is the following:
 \beqa
\frac{i}{2}(2\pi\alpha')(\alpha')^2\mu_p\left(\frac{\pi^2}{6}-8\ln(2)^2\right) C_{p-3}\wedge (F^{(1)}+F^{(2)})\wedge D ^aD_a(DT\wedge DT^*)\labell{hderv1}
\eeqa
This higher derivative term in field theory reproduces $exactly$ the contact term in the second line of \reef{line}.
The last term in \reef{line}, on the other  hand, is given by the  Feynman amplitude \reef{amp3} where the vertex $V_a(C_{p-3},A^{(1)},A^{(1)})$ should be derived from the higher derivative term \reef{highaa}, that is 
\beqa
V_a(C_{p-3},A^{(1)},A^{(1)})&=&\frac{(2\pi)^2}{6}\mu_p(2\pi\alpha')^2\frac{1}{(p-2)!}\epsilon_{a_0\cdots a_{p-1}a}H^{a_0\cdots a_{p-3}}k_1^{a_{p-2}}\xi^{a_{p-1}}(k_1\cdot p)^2
\eeqa
Note that the vertex $V_b(A^{(1)},T_1,T_2)$ has no higher derivative correction as it arises from the kinetic term of the tachyon. The tachyon pole of string amplitude indicates that the kinetic term has no higher derivative correction. Now replacing $V_b(A^{(1)},T_1,T_2)$ and the above vertex in \reef{amp3} one again finds exact agreement with the string theory amplitude. Hence the expansion of the string amplitude \reef{pn2} around \reef{point} produces an expansion whose  leading order terms are reproduced  by the Feynman amplitudes resulting fron DBI and WZ actions and the next  terms are related to the higher derivative corrections to the WZ action. In particular, its massless poles gives information about the higher derivative corrections of $C_{p-3}\wedge F\wedge F$ and its contact terms about $C_{p-3}\wedge F\wedge DT\wedge DT^*$.

\subsection{$p=n$ case}

Now we consider  $n=p$ case. The string theory amplitude in this case is symmetric under interchanging $2 \leftrightarrow 3$. On the other hand, there is no Feymann amplitude in field theory corresponding to four-point function of one RR, one gauge field, one $T_1$ and one $T_2$. Hence, for $p=n$ the string theory amplitude \reef{gen} is the S-matrix element of one RR, one gauge field and two $T_1$ or two $T_2$. Its electric part is,
\beqa
{\cal A}^{AT_1T_1C}&=&\pm\frac{8i\mu_p}{\sqrt{2\pi}p!}\left[\bigg( \eps^{a_{0}\cdots a_{p-1}a}H_{a_{0}\cdots a_{p-1}}\bigg)J\right. \nonumber\\&&\left.\times
\bigg\{k_{2a}(t+1/4)(2\xi.k_{3})
+k_{3a}(s+1/4)(2\xi.k_{2})-\xi_a(s+1/4)(t+1/4)\bigg\}\right]\labell{pn}\eeqa
 Note that the amplitude satisfies the Ward identity, \ie the amplitude vanishes under replacement $\xi^a\rightarrow k_1^a$. 
\begin{center}
\begin{picture}
(600,100)(0,0)
\Line(25,105)(75,70)\Text(50,105)[]{$T_{1}$}
\Line(25,70)(75,70)\Text(45,80)[]{$T_{1}$}
\Photon(25,35)(75,70){4}{7.5}\Text(50,39)[]{$A^{(1)}$}
\Photon(75,70)(125,70){4}{7.5}\Text(105,88)[]{$A$}
\Gluon(125,70)(175,105){4.20}{5}\Text(145,105)[]{$C_{p-1}$}
\SetColor{Black}
\Vertex(75,70){1.5} \Vertex(125,70){1.5}
\Text(105,20)[]{(a)}
\Line(295,105)(345,70)\Text(320,105)[]{$T_{1}$}
\Line(295,70)(345,70)\Text(320,80)[]{$T_1$}
\Photon(295,35)(345,70){4}{7.5}\Text(320,35)[]{$A^{(1)}$}
\Gluon(345,70)(395,105){4.20}{5}\Text(365,105)[]{$C_{p-1}$}
\Vertex(345,70){1.5}
\Text(345,20)[]{(b)}

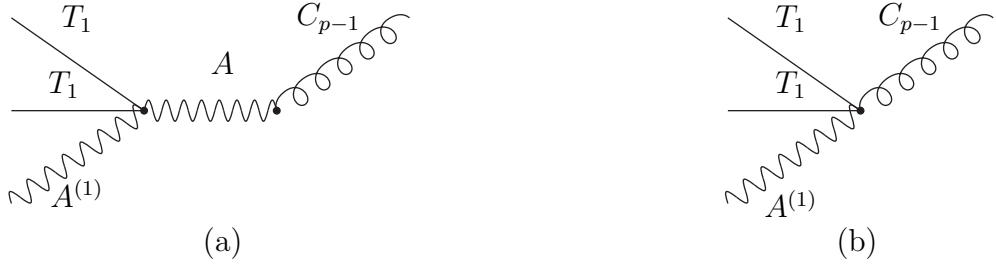
\end{picture}\\ {\sl Figure 3 : The Feynman diagrams corresponding to the amplitudes in \reef{amp4} and \reef{amp61}.} 
\end{center}
The effective field theory, \reef{exp1} and \reef{exp2}, has the following massless pole for $p=n$:
\beqa
{\cal A}&=&V_a(C_{p-1},A)G_{ab}(A)V_b(A,T_1,T_1,A^{(1)})\labell{amp4}\eeqa
where $A$ should  be $A^{(1)}$ and $A^{(2)}$. The propagator and vertexes  $ V_a(C_{p-1},A)$ and $V_b(A,T_1,T_1,A^{(1)})$ are
\beqa
G_{ab}(A) &=&\frac{i\delta_{ab}}{(2\pi\alpha')^2 T_p
\left(u+t+s+1/2\right)}\nonumber\\
V_a(C_{p-1},A^{(1)})&=&i\mu_p(2\pi\alpha')\frac{1}{p!}\epsilon_{a_0\cdots a_{p-1}a}H^{a_0\cdots a_{p-1}}\nonumber\\
V_a(C_{p-1},A^{(2)})&=&-i\mu_p(2\pi\alpha')\frac{1}{p!}\epsilon_{a_0\cdots a_{p-1}a}H^{a_0\cdots a_{p-1}}\nonumber\\
V_b(A^{(1)},T_1,T_1,A^{(1)})&=&-2iT_p(2\pi\alpha')\xi_b\nonumber\\
V_b(A^{(2)},T_1,T_1,A^{(1)})&=&2iT_p(2\pi\alpha')\xi_b\labell{Fey}
\eeqa
The amplitude \reef{amp4} then becomes
\beqa
{\cal A}&=&\frac{4i\mu_p}{p!(u+s+t+1/2)}\eps^{a_{0}\cdots a_{p-1}a}H_{a_{0}\cdots a_{p-1}}\xi_a\labell{amp6}\eeqa
The couplings in \reef{contact} has also  the following contact term:
\beqa
{\cal A}_c&=&i\mu_p\ln(2)\frac{16}{p!} \eps^{a_{0}\cdots a_{p-1}a}H_{a_{0}\cdots a_{p-1}}\xi_a\labell{amp61}\eeqa
Now in  string theory side, expansion of $(s+1/4)(t+1/4)J$ in \reef{pn} around \reef{point} is
\beqa
(s+1/4)(t+1/4)J&\!\!\!\!\!=\!\!\!\!\!&\frac{\sqrt{2\pi}}{2}\left(\frac{-1}{(t+s+u+1/2)}+ 4\ln(2)\right.\labell{high}\\
&&\left.+\left(\frac{\pi^2}{6}-8\ln(2)^2\right)(s+t+u+1/2)-\frac{\pi^2}{3}\frac{(t+1/4)(s+1/4)}{(t+s+u+1/2)}+\cdots\right)\nonumber\eeqa
Replacing it in \reef{pn}, one finds that the first term above is reproduced by the field theory massless pole \reef{amp6} and the second term by \reef{amp61}. To find the higher derivative term corresponding to the first term in the second line above, we note that the WZ terms in the third line of \reef{exp2} indicates that the following higher derivative term should accompany the term in \reef{hderv}: 
\beqa
-(\alpha')^2\mu_p\left(\frac{\pi^2}{6}-8\ln(2)^2\right) C_{p-1}\wedge D^aD_a[(F^{(1)}-F^{(2)})|T|^2]\labell{hderv2}
\eeqa
Combining  the above with  the  coupling of one RR, two tachyons and one gauge field  of \reef{hderv}, one finds the following coupling:
\beqa
-(\alpha')^2\mu_p\left(\frac{\pi^2}{6}-8\ln(2)^2\right) H_p\wedge \partial^a\partial_a[(A^{(1)}-A^{(2)})TT^*]\eeqa
which reproduce exactly the first term in the second line of \reef{high}. 

\begin{center}
\begin{picture}
(600,100)(0,0)
\Line(25,105)(75,70)\Text(50,105)[]{$T_{1}$}
\Line(75,70)(125,70)\Text(100,80)[]{$T_{2}$}
\Line(125,115)(125,70)\Text(120,95)[]{$T_{1}$}
\Photon(25,35)(75,70){4}{7.5}\Text(50,40)[]{$A^{(1)}$}
\Photon(125,70)(175,70){4}{7.5}\Text(150,80)[]{$A$}
\Gluon(175,70)(225,105){4.20}{5}\Text(200,110)[]{$C_{p-1}$}
\SetColor{Black}
\Vertex(75,70){1.5} \Vertex(125,70){1.5}\Vertex(175,70){1.5}
\Text(125,20)[]{(a)}
\Line(275,105)(325,70)\Text(300,105)[]{$T_{1}$}
\Line(325,70)(375,70)\Text(350,80)[]{$T_2$}
\Photon(275,35)(325,70){4}{7.5}\Text(300,35)[]{$A^{(1)}$}
\Line(375,70)(425,35)\Text(400,45)[]{$T_{1}$}
\Gluon(375,70)(425,105){4.20}{5}\Text(400,110)[]{$C_{p-1}$}
\Vertex(325,70){1.5}\Vertex(375,70){1.5}
\Text(350,20)[]{(b)}

\end{picture}\\ {\sl Figure 4 : The Feynman diagrams corresponding to the amplitudes \reef{amp5}.} 
\end{center}
The effective field theory has also the following  poles for $p=n$:
\beqa
{\cal A}&=&V_a(C_{p-1},A)G_{ab}(A)V_b(A,T_1,T_2)G(T_2)V(T_2,T_1,A^{(1)})\nonumber\\
&&+V(C_{p-1},T_1,T_2)G(T_2)V(T_2,T_1,A^{(1)})\labell{amp5}\eeqa
where the off-shell gauge field $A$ should be $A^{(1)}$ and $A^{(2)}$, and 
\beqa
V(C_{p-1},T_1,T_2)&=&\beta^2\mu_p(2\pi\alpha')\frac{1}{p!} \eps^{a_{0}\cdots a_{p-1}a}H_{a_{0}\cdots a_{p-1}}k_{2a}\nonumber\\
V(T_2,T_1,A^{(1)})&=&T_p(2\pi\alpha')(k_3-k)\cdot\xi\nonumber\\
V_b(A,T_1,T_2)&=&T_p(2\pi\alpha')(k_{2a}+k_a)\nonumber\\
G(T_2)&=&\frac{i}{(2\pi\alpha')T_p(s+1/4)}\labell{ver2}\eeqa
where $k_a$ is momentum of the off-shell tachyon. Replacing them in \reef{amp5}, one finds
\beqa
4\mu_p\left(\frac{-1}{(s+1/4)(t+s+u+1/2)}+ \frac{4\ln(2)}{(s+1/4)}\right)\frac{1}{p!} \eps^{a_{0}\cdots a_{p-1}a}H_{a_{0}\cdots a_{p-1}}k_{2a} (2k_3\cdot\xi)\labell{amp51}\eeqa
Similar result appear for the Feynmann diagram in which $k_2\leftrightarrow k_3$.
Now in string theory side, expansion of $(t+1/4)J$ around \reef{point} is
\beqa
(t+1/4)J&=&\frac{1}{2}\sqrt{2\pi}\left(\frac{-1}{(s+1/4)(t+s+u+1/2)}+ \frac{4\ln(2)}{(s+1/4)}\right.\labell{high2}\\
&&\left.+\left(\frac{\pi^2}{6}-8\ln(2)^2\right)\frac{(s+t+u+1/2)}{(s+1/4)}-\frac{\pi^2}{3}\frac{(t+1/4)}{(t+s+u+1/2)}+\cdots\right)\nonumber\eeqa
Replacing it in the string theory amplitude \reef{pn},   one finds exact agreement between the first two terms and the field theory results \reef{amp51}. The first term in the second line should be reproduced by the following Feynman amplitude in field theory:
\beqa 
{\cal A}&=&V(C_{p-1},T_1,T_2)G(T_2)V(T_2,T_1,A^{(1)})\labell{finalA}
\eeqa
where the propagator and the vertex $V(T_2,T_1,A^{(1)})$ is the same as in \reef{ver2}, however, the vertex $V(C_{p-1},T_1,T_2)$ should be derived from the higher derivative term \reef{hderv}, that is
\beqa
V(C_{p-1},T_1,T_2)&=&(\alpha')^2\mu_p\left(\frac{\pi^2}{6}-8\ln(2)^2\right)\frac{1}{p!} \eps^{a_{0}\cdots a_{p-1}a}H_{a_{0}\cdots a_{p-1}}k_{2a}(s+t+u+1/2)\nonumber\eeqa
Replacing them in \reef{finalA}, one finds exact agreement with the string theory result. Finally, the sum of last term in  \reef{high2} and the corresponding term when $k_2\leftrightarrow k_3$, and the last term in \reef{high} is
\beqa
-\frac{4i}{3}\pi^2\mu_p\frac{ \eps^{a_{0}\cdots a_{p-1}a}H_{a_{0}\cdots a_{p-1}}}{p!(s+t+u+1/2)}
\left[k_{2a}(t+1/4)(2\xi.k_{3})
-
\frac{1}{2}\xi_a(s+1/4)(t+1/4)+(1\leftrightarrow 2)\right]\nonumber\eeqa
To find the corresponding term in field theory, consider the following Feynman amplitude: 
\beqa
{\cal A}&=&V_a(C_{p-1},A)G_{ab}(A)V_b(A,A^{(1)},T_1,T_1)\labell{finalamp}\eeqa
where the vertex $V_a(C_{p-1},A)$ is the same as the one appears in \reef{Fey} and $ V_b(A,A^{(1)},T_1,T_2)$ is derived from the second line of \reef{exp1}, that is
\beqa
V_b(A^{(1)},A^{(1)},T_1,T_1)&\!\!\!=\!\!\!&\frac{4i}{3}T_p(\pi\alpha')^3k_b\left[(s+1/4)(2k_2\cdot\xi)+(t+1/4)(2k_3\cdot\xi)\right]+2iT_p(\pi\alpha')^3\times\nonumber\\
&&\times\left[k_{2b}(t+1/4)(2\xi.k_{3})
+k_{3b}(s+1/4)(2\xi.k_{2})-\xi_b(s+1/4)(t+1/4)\right]\nonumber\\
V_b(A^{(2)},A^{(1)},T_1,T_2)&\!\!\!=\!\!\!&\frac{2i}{3}T_p(\pi\alpha')^3k_b\left[(s+1/4)(2k_2\cdot\xi)+(t+1/4)(2k_3\cdot\xi)\right]+2iT_p(\pi\alpha')^3\times\nonumber\\
&&\times\left[k_{2b}(t+1/4)(2\xi.k_{3})
+k_{3b}(s+1/4)(2\xi.k_{2})-\xi_b(s+1/4)(t+1/4)\right]\nonumber
\eeqa
where $k^a$ is the momentum of the off-shell gauge field. Replacing them in \reef{finalamp}, one finds exact agreement with the string theory result. Note that the $F^{(1)}\cdot F^{(2)}$ term in the tachyon DBI action \reef{exp1} is necessary for the above consistency. The above consistency indicates that the coupling $C_{p-1}\wedge F$ has no higher derivative correction. 

Now let us speculate on the other terms of the string theory expansion. Since there is no higher derivative corrections to $C_{p-1}\wedge F$ and to the kinetic term, the expansion should not have the double poles other than the one appears in the leading term. For the simple tachyonic poles, since the kinetic term has no correction, the non-leading poles of tachyon should gives information about the higher derivative corrections to $C_{p-1}\wedge DT\wedge DT^*$. For the simple massless poles, since there is no correction to  $C_{p-1}\wedge F$, the non-leading massless poles should give information about the higher derivative corrections to the coupling of two tachyons and two gauge fields.  Finally, the contact terms of the string theory amplitude gives information about the higher derivative correction to  $C_{p-1}\wedge DT\wedge DT^*$ and to $C_{p-1}\wedge F TT^*$. It would be interesting  to study  these terms in details and find higher derivative corrections to tachyon DBI and to WZ actions.
\section*{Acknowledgment}

We would like to thank A. Ghodsi for  comments .


\end{document}